\documentclass[letterpaper]{article} 
\usepackage{aaai25}  
\usepackage{times}  
\usepackage{helvet}  
\usepackage{courier}  
\usepackage[hyphens]{url}  
\usepackage{graphicx} 
\usepackage{natbib}  
\usepackage{caption} 
\usepackage{longtable}
\usepackage{algorithm}
\usepackage{algorithmic}

\usepackage[x11names]{xcolor}
\usepackage{color-edits} 
\addauthor{ic}{Orchid3}

\usepackage{booktabs}

\usepackage{pdflscape}

\usepackage{newfloat}
\usepackage{listings}
\DeclareCaptionStyle{ruled}{labelfont=normalfont,labelsep=colon,strut=off} 
\floatstyle{ruled}
\newfloat{listing}{tb}{lst}{}
\floatname{listing}{Listing}
\title{Bridging Research Gaps Between Academic Research and \\Legal Investigations of Algorithmic Discrimination}
\author{
    Colleen V. Chien\textsuperscript{\rm 1}, Anna Zink\textsuperscript{\rm 2}, Irene Y. Chen\textsuperscript{\rm 1,3}
}
\affiliations{
    \textsuperscript{\rm 1}University of California, Berkeley\\
    \textsuperscript{\rm 2}Tufts University\\
    \textsuperscript{\rm 3}University of California, San Francisco\\

}

\usepackage{bibentry}

\begin{document}

\maketitle

\begin{abstract}
As algorithms increasingly take on critical roles in high-stakes areas such as credit scoring, housing, and employment, civil enforcement actions have emerged as a powerful tool for countering potential discrimination. These legal actions increasingly draw on algorithmic fairness research to inform questions such as how to define and detect algorithmic discrimination. However, current algorithmic fairness research, while theoretically rigorous, often fails to address the practical needs of legal investigations. We identify and analyze 15 civil enforcement actions in the United States including regulatory enforcement, class action litigation, and individual lawsuits to identify practical challenges in algorithmic discrimination cases that machine learning research can help address. Our analysis reveals five key research gaps within existing algorithmic bias research, presenting practical opportunities for more aligned research: 1) finding an equally accurate and less discriminatory algorithm, 2) cascading algorithmic bias, 3) quantifying disparate impact, 4) navigating information barriers, and 5) handling missing protected group information. We provide specific recommendations for developing tools and methodologies that can strengthen legal action against unfair algorithms. 

\end{abstract}

%

\section{Introduction}
\label{sec:intro}
In June 2022, the US Department of Justice announced a settlement agreement with Meta regarding allegedly discriminatory housing advertisement algorithms on Facebook~\cite{meta2024}. The settlement required Meta to stop using protected characteristics such as race and gender in their ad targeting systems. It mandated the development of non-discriminatory alternatives, and imposed a \$115,054 civil penalty, the maximum allowed under the Fair Housing Act.  This case represents a broader trend over the last decade of  enforcement agencies and private plaintiffs turning to  regulatory action and litigation to address algorithmic discrimination. As algorithmic decision-making systems increasingly shape critical outcomes from housing to healthcare, legal protections represent one of the most important bulwarks against algorithmic harm---but only to the extent they are enforceable.

The legal framework for addressing algorithmic discrimination in the US is built upon the 5th and 14th Amendments to the Constitution, which provide due process and equal protection safeguards against state action, and a body of federal civil rights laws extending these protections to private action. These laws, including the Civil Rights Act of 1964 and the Fair Housing Act of 1968, create mechanisms to challenge discrimination in areas like employment, housing, and public accommodations. Courts and regulators have applied existing authorities to emerging algorithmic systems, resulting in enforcement actions and court decisions (e.g. such as~\cite{mobley2024workday}). States and localities have also started to enact new laws to regulate AI, such as the Colorado Artificial Intelligence Act of 2024, which creates a duty of reasonable care to guard against known or reasonably foreseeable risks of algorithmic discrimination ~\cite{coloradoConsumerProtections}. Illinois' HB3773 protects employees from AI-related discrimination in employment contexts~\cite{ilgaIllinoisGeneral},
following on city-level efforts such as New York City's Local Law 144~\cite{wright2024null}. Such laws have been motivated in part by reports across multiple sectors demonstrating how automated decision-making can perpetuate discrimination through, e.g., hiring algorithms that systematically downgrade qualified female candidates~\cite{dastin2018}, tenant screening systems that disproportionately reject minority applicants~\cite{thevergeLandlordScreening}, and healthcare allocation algorithms that provide lower levels of rehabilitation care to elderly patients~\cite{statnewsUnitedHealthFaces}. Automated decision-making can perpetuate or amplify historical patterns of discrimination, even without explicit intent.

Concerns about algorithmic discrimination have motivated the machine learning (ML) and related communities to focus their efforts on three main goals: the formalization of fairness metrics~\cite{barocas2023fairness,chouldechova2017fair,heidari2018fairness,corbett2023measure}, the development of fairness-aware ML~\cite{dwork2012fairness,hardt2016equality,kusner2017counterfactual,friedler2021possibility}, and the creation of algorithmic auditing frameworks~\cite{chen2018my,buolamwini2018gender,raji2020closing}. However, there remains a large gap between the academic literature on algorithmic fairness and the practical tools needed by regulators and plaintiffs to evaluate algorithms for potentially actionable discrimination.  Critics note that fairness metrics often fail to align with legal standards~\cite{xiang2020reconciling,ho2020affirmative,wachter2020bias,watkins2024four}, highlighting the need for the ML community to align its research agendas to better serve interdisciplinary stakeholders ~\cite{henderson2024rethinking}.

\begin{figure*}
    \centering
    \includegraphics[width=0.55\linewidth]{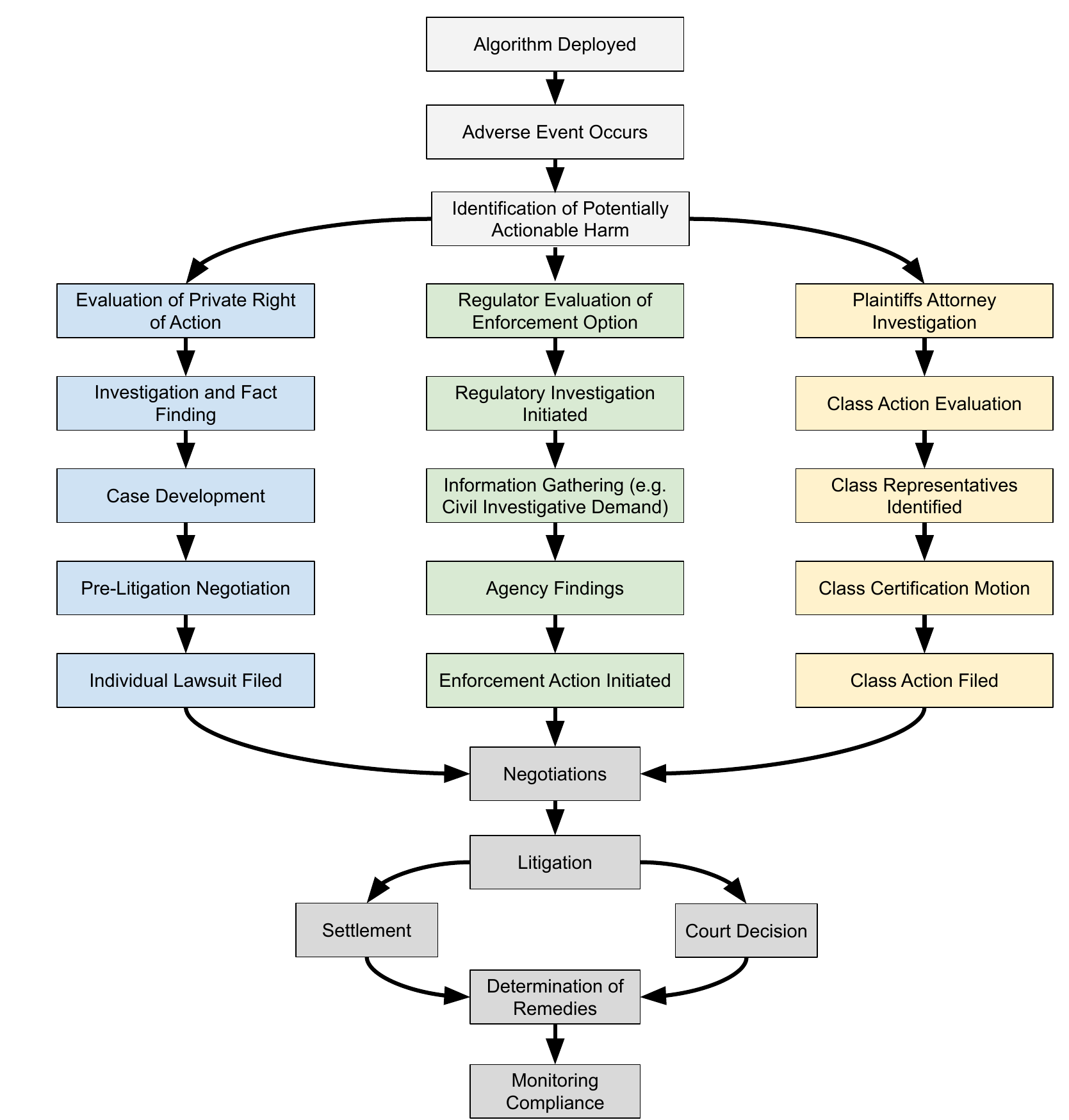}
    \caption{Flowchart of different types of civil enforcement actions.}
    \label{fig:v2_flowchart}
\end{figure*}

This work analyzes the disconnect between existing research and legal enforcement through the lens of existing legal actions targeting algorithmic discrimination. 
As shown in Figure~\ref{fig:v2_flowchart}, legal actions to regulate algorithms can take several forms, including individual lawsuits filed on behalf of named plaintiffs, class action lawsuits, and  enforcement actions taken by or urged of regulators such as state attorney generals or federal enforcers. Our analysis demonstrates the need for greater alignment between academic research and legal efforts to strengthen enforcement and promote ethical algorithmic development.

Our contribution is two-fold. First, we identify and analyze 15 civil enforcement actions targeting algorithmic discrimination across sectors including insurance, employment, and housing. We surface critical patterns in algorithmic bias investigations by drawing on the legal arguments and allegations contained in complaints, the legal standards for demonstrating actionable algorithmic discrimination, and various remediation approaches mandated by settlements. Second, we synthesize our findings into five concrete research directions to bridge the gap between academic work and legal practice: (1) finding an equally accurate and less discriminatory algorithm, (2) cascading algorithmic bias, (3) quantifying disparate impact, (4) information barriers in algorithmic investigations, and (5) missing protected group information.

\section{Related Work}
\label{sec:related}

\subsection{Definitions of Algorithmic Discrimination}

\paragraph{Algorithmic Fairness} In machine learning communities, discrimination has often been formalized as ``algorithmic fairness''~\cite{barocas2023fairness}, generally referring to a difference between pre-specified sensitive attributes for a computable metric of interest. Examples of these metrics include accuracy, calibration, or false positive rates~\cite{chouldechova2017fair,chen2019can,kleinberg2016inherent,seyyed2021underdiagnosis}. A large branch of research has focused on learning ``fairness-aware'' algorithms by optimally adjusting a learned prediction model to remove discrimination, mathematically formalized through a predetermined and computable metric. Several methods have focused on adjusting the algorithm itself~\cite{hardt2016equality,kamishima2011fairness,kamiran2010discrimination,zafar2017fairness,zemel2013learning} while others have focused on the underlying data~\cite{hajian2013methodology,feldman2015certifying,calmon2017optimized,chen2020ethical,shen2024data}. Through these efforts, the notion of a fairness-accuracy tradeoff has emerged, with post hoc correction methods available to navigate the balance~\cite{woodworth2017learning,hardt2016equality}. The community appears to have coalesced on several commonly-accepted algorithmic fairness metrics and mitigation methods, leading researchers to implement these calculations in publicly-available software packages~\cite{arya2021ai,saleiro2018aequitas,pfohl2024toolbox}.



\paragraph{Disparate Treatment and Disparate Impact} 
In contrast with the emphasis of machine learning research on model performance metrics and outcome metrics,   
legal compliance across domains requires specific contextual evaluation.~\cite{jacobs2000fair}. Algorithmic discrimination doctrine centers on two theories: disparate treatment and disparate impact. Disparate treatment involves intentional discrimination on the basis of protected characteristics, for example,  algorithms that explicitly include the attributes or their proxies and differ in their treatment of groups of individuals on the basis of these attributes or proxies~\cite{barocas2023fairness}. Disparate impact
theories implicate practices that disproportionately harm protected groups without requiring proof of discriminatory intent. The algorithmic fairness literature often simplifies these concepts, equating disparate treatment with the use of protected variables~\cite{xiang2019legal,dwork2012fairness,lipton2018does} and disparate impact with disproportionate outcomes~\cite{xiang2019legal, corbett2023measure, feldman2015certifying}. However, scholars have noted fundamental tensions between technical fairness metrics and legal discrimination frameworks~\cite{xiang2020reconciling,xiang2019legal,wachter2020bias}.

\paragraph{Other Legal Definitions} \label{sec:four-fifths} Although disparate treatment and disparate impact provide the main legal frameworks for analyzing algorithmic discrimination, other relevant legal concepts are worth noting. One example is the four-fifths rule (also known as the 4/5 rule or 80\% rule).  This general rule of thumb, as promulgated by the Guidelines of the US Equal Employment Opportunity Commission (EEOC), is a method for determining whether the selection rate for one group is ``substantially'' different than the selection rate of another group, potentially evincing disparate impact-based discrimination~\cite{eeocSelectIssues}.  The rule states that one rate is substantially different than another if their ratio is less than four-fifths (or 80\%) and has been endorsed by the Supreme Court in Griggs v. Duke Power Co.~\cite{griggs1971}. Despite being one component of a broader legal framework, and despite the EEOC itself  indicating that the four-fifths rule is ``merely a rule of thumb''  that may be ``inappropriate under certain circumstances,''  it has been disproportionately emphasized in ML literature, often misrepresenting actual disparate impact doctrine~\cite{watkins2024four}.

The legal discourse around algorithmic discrimination is further shaped by two fundamental principles: anti-classification and anti-subordination~\cite{balkin2003american}. Anti-classification holds that any differential treatment based on protected attributes is discriminatory, often interpreted as prohibiting protected class-conscious decision-making. In contrast, anti-subordination argues that law should actively work to dismantle group-based hierarchies, even if this requires explicit consideration of protected class status. As a field, ML has focused primarily on anti-classification while neglecting anti-subordination~\cite{xiang2019legal,xiang2020reconciling}.

\subsection{Algorithmic Discrimination Oversight}
Algorithmic discrimination oversight in the US operates through multiple legal channels as shown in Figure 1. While shown separately, in practice, these pathways may intersect, for example, with the evaluation of a private right of action or class action becoming the basis for an agency enforcement action. Federal agencies like the Federal Trade Commission (FTC), EEOC, and Consumer Financial Protection Bureau (CFPB) have historically taken a leading role in enforcing anti-discrimination protections. In recent years, the FTC has taken action against Rite Aid's facial recognition safeguards~\cite{ftc_v_riteaid_2023_complaint}, and been urged to initiate action against HireVue's AI assessment tools~\cite{epic2019hirevue}, and Aon's employment screening technology~\cite{aclu2024aon} while the EEOC has provided guidance on AI hiring tools and investigated automated discrimination~\cite{eeocSelectIssues}.
Beyond agency enforcement, private litigation through class action suits and individual claims (Figure~\ref{fig:v2_flowchart}) establish precedents for applying discrimination frameworks to algorithms, exemplified by Mobley v. Facebook and Mobley v. Workday~\cite{mobley2016, mobley2024workday}. Civil society organizations drive change through strategic litigation, as seen in the settlement between National Fair Housing Alliance (NFHA) and Facebook~\cite{nfha_v_facebook_2018} and the American Civil Liberties Union's credit scoring algorithm challenges~\cite{sandvig_v_lynch_2016}. Several of these actions are described in Table 1 and were analyzed as part of this project.

While federal and private actors shape enforcement, state and local governments have enacted their own algorithmic accountability legislation. New York City's Local Law 144 requires employment decision tool audits~\cite{groves2024auditing}, 
Illinois mandates AI interview disclosures~\cite{il-ai-video-2020}, 
and, starting in 2026, Colorado will require developers and deployers of certain AI systems to use "reasonable care" to protect consumers from the risks of algorithmic discrimination ~\cite{coloradoConsumerProtections}. However, similar efforts have faced resistance in other states~\cite{apnewsAttemptsRegulate}, and by the Trump White House ~\cite{TrumpEO}.
Regulatory approaches, including agency rulemaking like the CFPB's automated credit guidance~\cite{consumerfinanceCFPBIssues}, industry self-regulation, academic partnerships, and international laws such as the AI Act of the European Union~\cite{artificialintelligenceactArtificialIntelligence} complement domestic law making and enforcement activities. 
Comprehensive surveys of regulatory measures emphasize the evolving nature of U.S. legal practices in addressing different types of algorithmic discrimination~\cite{wang2024algorithmic}.

\subsection{Academic-Legal Fairness Gaps}



Scholars have identified a fundamental disconnect between academic algorithmic fairness research and legal discrimination investigations in how they define and measure bias. Current algorithmic fairness definitions often fail to capture the contextual and historical dimensions central to civil rights law~\cite{green2022flaws,watkins2024four}. Researchers have identified five key pitfalls in technical approaches, including the ``framing trap'' of failing to model the entire system, the ``portability trap'' of failing to understand the risks of repurposing algorithm for a different context, and the ``formalism trap'' of reducing discrimination to purely mathematical terms~\cite{selbst2019fairness}. 
The academic focus on individual fairness metrics further conflicts with civil rights law's emphasis on group-based disparate impact analysis~\cite{mulligan2019thing}. Recent work has called for rethinking ML benchmarks to better align with professional codes of conduct and legal standards~\cite{henderson2024rethinking}.

These theoretical challenges manifest in practical implementation barriers. Enforcement agencies struggle to apply academic fairness tools in real investigations, facing obstacles in data access, computational resources, and developing legally defensible methodologies~\cite{raji2020closing,metcalf2021algorithmic}. This implementation gap has sparked debate about race-aware systems, with some arguing they constitute impermissible disparate treatment~\cite{bent2019algorithmic,barocas2016big,kroll165accountable}, while others uphold them as acceptable with proper safeguards~\cite{kim2022race,hellman2020measuring}.

A fundamental limitation underlies both theoretical and practical challenges: current fairness metrics inadequately address causality, particularly in court-mandated algorithmic remediation~\cite{xiang2020reconciling,xiang2019legal, hellman2020measuring}. This gap in understanding how different interventions affect disparate impact significantly hampers efforts to address algorithmic discrimination effectively.


\section{Methods}
\label{sec:methods}


To carry out the analysis reported in this article, we first searched for civil enforcement actions involving well-pled claims of discrimination, locating 15 actions. We then proceeded to analyze these actions for gaps that could potentially be addressed by the research community.

\subsection{Identifying Civil Actions}
We identified civil enforcement actions, including civil lawsuits, settlements, complaints filed with regulators, and regulatory actions alleging algorithmic discrimination using a set of both targeted and broad searches. 

To locate existing cases, we entered into the Westlaw/Lexis advanced search engines---which cover all published appellate court opinions, some trial level and administrative actions, and some federal and state actions---the following search string as well as a variant that include the names of the agencies: \texttt{"CFPB" or "FTC" or "DOJ" or "attorney general" or "HUD" or "FHA" or "EEOC" or "SEC"): ATLEAST3("algorithm!") /s ("discrimin!" or "bias") and ("enforcement actions" or "settlement" or "litigation")}. This search was current as of May 16, 2025. 

This search yielded an initial result of 88 non-unique results across the two platforms, each of which was reviewed for relevance. This led to the exclusion of cases in which keyword mentions were incidental (such as the use of the term “discriminate” to mean to discern), the algorithm was tangential to the discrimination claim, or a claim of discrimination against a protected class was not present. However, we left in actions based not only on traditional discrimination law (e.g. asserted a violation of equal protection) but also on theories of unfair competition or consumer protection that included allegations of discriminatory harm. 

Because legal search engines routinely miss unpublished, trial, or lower court case proceedings, particularly at the state level, as well as records of agency proceedings, in addition to this “targeted” search of court proceedings, we also looked for relevant actions using a broad list of terms (see Table~\ref{tab:terms} in the appendix) using various search engines, including HeinOnline, Google and Google domain search (\texttt{site:.gov}), some of which were time bound, for example, involving a deep search of actions of the Federal Trade Commission that took place from 2015 to 2025 involving allegations related to algorithmic discrimination, algorithmic bias, algorithmic fairness, or deceptive practices. 

Our search was not limited to any particular industry, and generated actions involving algorithms from multiple domains such as housing, employment, and criminal risk assessment, in both public and private sectors. We narrowed our focus to investigations of real-world algorithms rather than guidance statements from regulatory bodies, federal or state legislature, or academic proposals for algorithmic auditing frameworks and included complaints filed with regulators by third parties alleging harm given that such complaints require the regulator to take further action. 

We chose cases that explicitly addressed discrimination against a protected class and left out actions where allegations of discrimination were merely asserted. This process resulted in a list of 15 civil enforcement actions across federal and state authorities, a subset of which are shown in Table 1. Additional columns such as details about the type of action, the venue, the year, and the algorithmic anti-discrimination authority relied upon are provided in Table 3 in the appendix. 


\subsection{Data Analysis}
We analyzed the collected civil enforcement actions and surfaced within them consistent themes, patterns, and connections to algorithmic bias research. Rather than cataloging all possibilities, we focused our efforts on uncovering the most salient  ways in which ML research could better address the practical challenges of algorithmic bias investigations.

Each enforcement action centers on allegations of algorithmic discrimination. These cases broadly fall into two categories: (1) actions involving algorithms  alleged to explicitly use protected characteristics, such as Apple Card's screening algorithm's alleged gender discrimination~\cite{applecardpressrelease} and Allstate's credit underwriting algorithm's alleged racial discrimination~\cite{dehoyos2007}, or (2) cases in which algorithms demonstrated performance issues or inadequate safeguards, as exemplified by Rite Aid's facial recognition system~\cite{ftc_v_riteaid_2024_order}. 

\clearpage

\clearpage
\onecolumn

\begin{small}
\begin{landscape}
    \begin{longtable}{p{5cm}p{4cm}p{10cm}p{2cm}}
\toprule
\textbf{Complainant(s)}                                                                                           & \textbf{Respondent(s)}                                                & \textbf{Algorithm and Harm Alleged}                                                                                                                                                                                                       & \textbf{Status as of May 2025}                                                                                                                            \\ \midrule 
Louis and other rental applicants in Massachusetts & SafeRent Solutions LLC & Algorithmic tenant screening program alleged to have disparate impact on low-income Black and Hispanic housing voucher recipients and housing applicants & Settlement reached \\
\midrule

US Federal Trade Commission & Rite Aid Corp. & Facial recognition technology alleged to erroneously identify shoplifters, with high false-positive matches especially likely among Black, Latino, Asian, and female consumers & Settlement reached \\
\midrule

US Equal Employment Opportunity Commission & iTutor & Automated system in recruiting software alleged to reject female applicants over 55 years old and male applicants over 60 years old & Settlement reached \\
\midrule

US Department of Justice & Meta & Ad-delivery system for housing advertisements alleged to discriminatorily target groups based on race, ethnicity, and sex & Settlement reached \\
\midrule

DeHoyos and other housing applicants in Texas and Florida & Allstate & Automated credit scoring algorithms alleged to have an adverse disparate impact on minority insurance applicants by charging them unfairly high premiums & Settlement reached \\
\midrule

Flores and other inmates in New York & Stanford and other members of New York State Board of Parole & Predictive risk assessment tool (COMPAS) alleged to calculate recidivism scores without factoring in an inmate's demonstrated maturity and rehabilitation, thus minimizing younger inmates' chances of release & Settlement reached \\
\midrule

Huskey and other homeowners in the Midwest & State Farm & Homeowner insurance claims-processing algorithms alleged to predict higher levels of fraud from Black policyholders and thus subjected them to greater scrutiny & Case is pending \\
\midrule

Mobley and other job applicants & Workday & Job application automated screening tool alleged to discriminate against job applicants on the basis of race, age, and/or disability & Case is pending \\
\midrule

Oliver and other mortgage applicants & Navy Federal Credit Union & Mortgage lending algorithm decisions alleged to discriminate based on race by denying plaintiffs a loan or offering ones on less favorable terms & Settlement reached \\
\midrule

Real Women in Trucking & Meta & Ad-delivery algorithm alleged to target job applicants based on gender and age & Complaint pending \\
\midrule


Connecticut Fair Housing Center and Carmen Arroyo & CoreLogic Rental Property Solution & Tenant screening algorithmic tool (CrimSAFE) alleged to discriminate against persons with a criminal record, disproportionately adversely impacting Black and Latino applicants & Appeal pending before the 2nd Circuit \\
\midrule

Liapes & Facebook & Online advertising algorithm alleged to discriminate against women in the showing of life insurance ads & Case closed \\
\midrule

New York State Dept of Financial Services & Apple & Credit-decision and credit-limit algorithms alleged to discriminate against women & No evidence of deliberate discrimination \\

\bottomrule \\ 
    \caption{Subset of identified algorithmic discrimination investigations. See Table~\ref{tab:extended} in appendix for full set of 15 investigations and additional columns, including a) Type of Action, Venue, and Year Action Initiated, and b) Algorithmic Anti-Discrimination Regulation/Authority.}
    \label{tab:investgations}
    \end{longtable}
    \end{landscape}
\end{small}

\clearpage
\twocolumn

The cases have varied outcomes: several have reached settlements~\cite{thevergeLandlordScreening,cohenmilsteinAIRelatedDiscrimination,ftc_v_riteaid_2024_order,meta2024,dehoyos2007}, some found no wrongdoing~\cite{applecardinvestigation}, and ongoing cases continue to shape evolving standards for algorithmic auditing and compliance~\cite{huskey2022complaint,lokken2024,mobley2024workday}.

The legal justifications and remediation approaches in these cases highlight critical areas where ML research can better support enforcement.
Cases frequently rely on anti-discrimination frameworks like the Fair Housing Act and Equal Credit Opportunity Act (ECOA) to establish protected characteristics and determine permissible algorithmic inputs ~\cite{meta2024,dehoyos2001,cohenmilsteinAIRelatedDiscrimination}. However, enforcers face several challenges when attempting to  quantify discrimination: while some cases successfully demonstrated disparate impact through statistical evidence~\cite{dehoyos2007}, others struggled with protected group imputation and data access barriers~\cite{statnewsUnitedHealthFaces,meta2024}. The remediation approaches ordered through settlements reveal the practical need for research on algorithm modification techniques that maintain accuracy while reducing bias. This is particularly evident in cases where companies were required to redesign their algorithms~\cite{cohenmilsteinAIRelatedDiscrimination}, highlighting the gap between theoretical fairness improvements and deployable solutions. 
Financial penalties and technology bans, while effective enforcement tools, underscore the necessity of scrupulous algorithmic investigations~\cite{meta2024,cohenmilsteinAIRelatedDiscrimination,ftc_v_riteaid_2024_order}. These enforcement patterns inform our identified research opportunities, particularly around developing standardized disparate impact measurements and techniques for finding equally accurate, less discriminatory alternatives.

\subsection{Limitations of Study}
\label{sec:limitations}

The 15 actions that are the subject of this study do not represent the exhaustive universe of algorithmic-bias enforcement activity. A comprehensive review remains unfeasible due to the fragmented nature of U.S. enforcement and litigation systems, which extend beyond any single research platform or public docket. For example, federal and state agency records systems are non-unified and often siloed, with each agency using its own tracking system and standards for what to make public. Similarly, state filings and dockets oftentimes don't make it into digital repositories. Because search syntax, coverage, and metadata standards differ across courts, agencies, and time periods, no realistic combination of keyword, agency-name, and domain searches can surface every investigation or lawsuit.




\section{Identified Research Gaps}
\label{sec:findings}

Our investigation into the civil enforcement actions related to algorithmic discrimination revealed several gaps in methodology and tools needed to identify, measure, and evaluate algorithmic discrimination. 

\begin{figure*}[ht]
    \centering
    \includegraphics[width=0.65\linewidth]{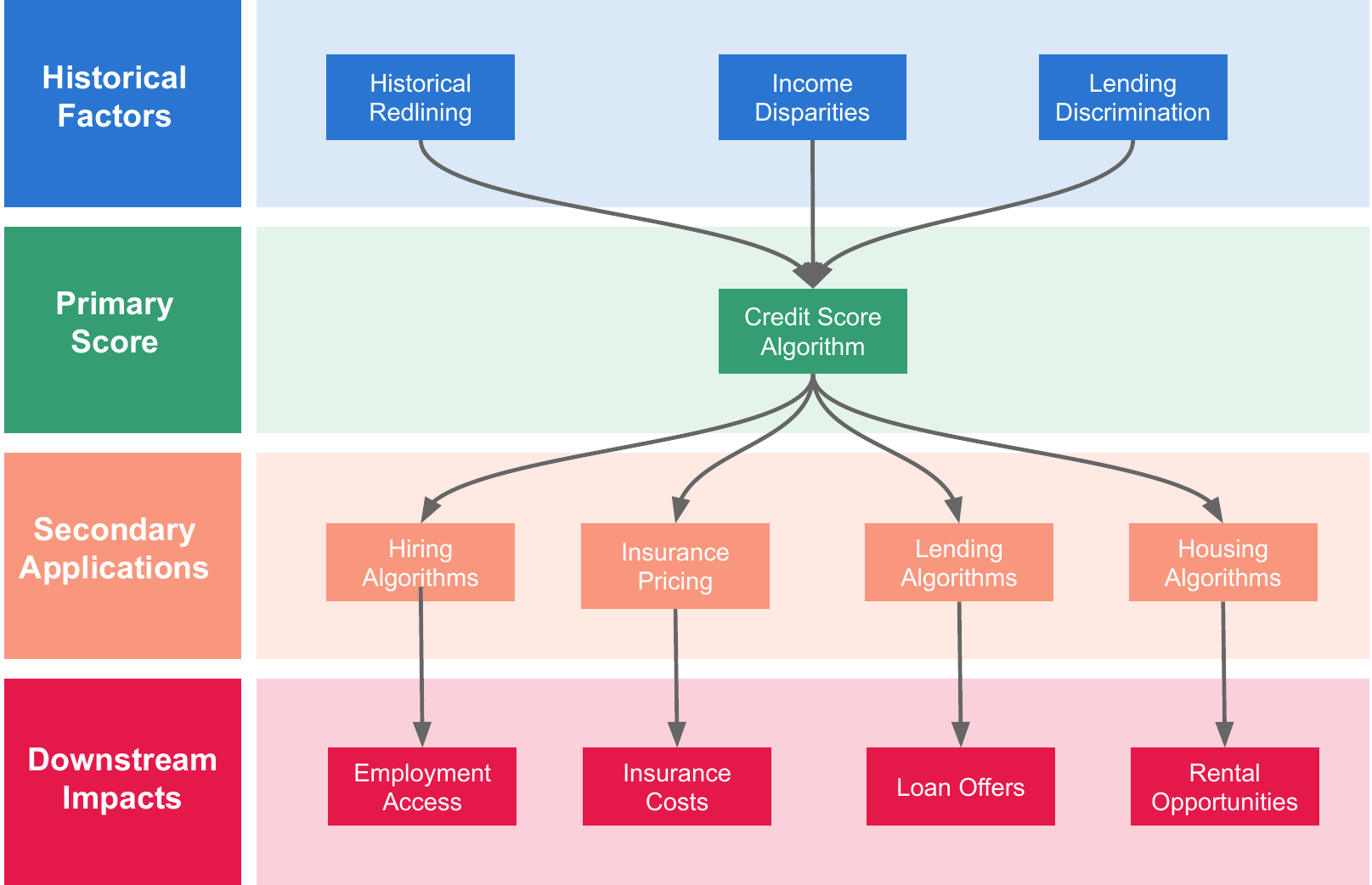}
    \caption{Illustration of how historical factors can affect a primary score (e.g., credit score), which then might be used for many secondary applications.}
    \label{fig:cascading}
\end{figure*}

\subsection{Finding an Equally Accurate and Less Discriminatory Algorithm}

Under US civil rights law, defendants in discrimination cases can justify practices with disparate impact if they demonstrate ``business necessity.'' However, this defense fails if plaintiffs can show the existence of less discriminatory alternatives that achieve similar business objectives. This concept has direct relevance to algorithmic discrimination cases. In 2022, a class action lawsuit of homeowners alleged that State Farm's homeowner insurance claims processing algorithms subjected Black policyholders to greater scrutiny. 
The complaint specifically questioned, \textit{``Whether substantially equally or more valid alternative means of claim processing are available that would eliminate or reduce the discriminatory impact [of State Farm's claims processing policy]''}~\cite{huskey2022complaint}. As of writing, the investigation remains ongoing as the plaintiffs have demonstrated a probable likelihood that their allegations have merit~\cite{huskey2023}. In 2007, a similar case emerged when plaintiffs sued Allstate for allegedly charging racially discriminatory insurance premiums. While denying these claims, Allstate agreed to implement a new pricing algorithm as part of the settlement~\cite{dehoyos2007}. Though the settlement documents do not describe the development of this alternative algorithm, it seems reasonable to assume that it was designed to balance non-discrimination with business performance requirements. Legal scholars have established that defendants have an affirmative duty to search for less discriminatory alternatives in many contexts~\cite{black2024legal}, and recent work has begun operationalizing this search in fair lending contexts~\cite{gillis2024operationalizing}.

This legal framework creates a critical technical challenge: identifying algorithms that maintain accuracy while reducing discriminatory impact. From a ML perspective, this relates to the ``Rashomon effect''. Named after the 1950 Japanese film about multiple narratives of the same event, the Rashomon effect describes the existence of multiple models with nearly equivalent accuracy but different internal structures~\cite{breiman2001statistical,fisher2019all}.  
Recent academic work has demonstrated the prevalence of this effect, showing that maximally accurate models can produce varying degrees of demographic disparities~\cite{rodolfa2020case}. This empirical evidence, combined with theoretical foundations, strongly suggests that less discriminatory alternatives could exist without sacrificing performance. 

However, a critical research gap severely hampers algorithmic discrimination investigations: we lack efficient methods to systematically explore and identify these equally accurate but less biased alternatives. This limitation has profound implications for civil rights litigation, where demonstrating the existence of less discriminatory alternatives is often legally required under disparate impact doctrine. Current approaches fall short in two ways: by either compromising the accuracy by training single "fair" models using constrained optimization~\cite{zafar2017fairness}, or they rely on computationally intensive searches through model spaces without guarantees of finding optimal trade-offs between accuracy and fairness metrics~\cite{kleinberg2016inherent}. The challenge is further exacerbated by theoretical impossibility results, which show that different fairness metrics often cannot be simultaneously satisfied~\cite{chouldechova2017fair}. This challenge is compounded by the need to consider algorithmic discrimination across different input types and accountability frameworks~\cite{bartlett2021algorithmic}.

The field needs research advances in three key areas to address these challenges. First, researchers should develop efficient search algorithms that can identify equally accurate but less discriminatory models within the Rashomon set, potentially drawing on techniques from multi-objective optimization and robust ML. Second, we need theoretical frameworks that can characterize the trade-off space between accuracy and various fairness metrics for different model classes, helping practitioners understand when less discriminatory alternatives are likely to exist. Finally, the field requires practical methodologies for comparing algorithmic alternatives that account for both immediate performance metrics and long-term societal impacts. These advances would directly support civil rights litigation by providing concrete tools for identifying and evaluating less discriminatory alternatives that maintain business performance.



\subsection{Cascading Algorithmic Bias}
\label{sec:cascading}

Legal investigations of algorithmic discrimination are complicated by the interconnected nature of modern scoring systems. When regulators or plaintiffs identify bias in an algorithm, they must trace discriminatory effects through multiple systems that may amplify the original disparity. This creates significant evidentiary challenges, as investigators must demonstrate not only the existence of bias in individual algorithms but also how these biases compound through automated decision chains. These complex chains, where outputs from one system become inputs to others can result in  ``cascading algorithmic bias'' as discrimination in one score percolates through downstream systems. For instance, as illustrated in Figure~\ref{fig:cascading}, a credit score influenced by historical factors such as redlining might be used in automated hiring, insurance pricing, and housing algorithms, further multiplying the original discriminatory effects~\cite{nelson2010credit}. Credit scores represent a well-documented example of cascading bias due to historical discrimination. Similarly, criminal background information is also used in a wide variety of contexts to screen individuals out of opportunities in non-criminal justice realms~\cite{umichquotAmericaapossPaper}.  


Cascading algorithmic bias is prominent in several of the cases we examined. For example, in a class action lawsuit filed by Massachusetts rental applicants against SafeRent LLC, plaintiffs alleged that the algorithmic tenant screening program disproportionately harmed housing voucher recipients. A key argument criticized the use of credit risk scores in SafeRent's resident screening risk score due to the historical context of credit scores: \textit{``Black, Hispanic, and low-income applicants and voucher holders are more likely to have poor credit histories, which means they are disproportionately likely to have lower credit scores''}~\cite{cohenmilsteinAIRelatedDiscrimination}. Notably, the complaint's argument did not rest solely on the use of credit scores, but instead alleged that individuals with federal and state housing voucher programs were improperly evaluated by the tenant screening algorithm, given that a housing voucher uniquely guarantees the housing provider’s receipt of monthly rent. 
The case was settled in 2024 with a \$2.28 million settlement and an agreement to modify the tenant screening algorithms, particularly regarding individuals with a publicly funded federal or state housing voucher in connection with the rental application.

While credit scores have been established as a known source of cascading bias, companies are increasingly selling proprietary scoring systems to other businesses that may perpetuate or amplify discrimination in novel ways. These custom algorithms often lack transparency and validation. 
For example, in an ongoing investigation against Workday -- a platform used by over 10,000 businesses to manage job applications -- plaintiffs claim that its algorithm disproportionately disqualifies job applicants on the bases of age, race, and disabilities \cite{mobley2024workday}. Finally, devastating effects of algorithmic errors emerged in a case brought against Rite Aid and its use of facial recognition technology for surveillance purposes. Rite Aid identified ``persons of interest'' in a database using low-resolution images and the technology would create an alert when a match was found. The technology has a high false positive rate especially among Black, Latino, Asian, and female consumers: \textit{``[D]uring a five-day period, Rite Aid’s facial recognition technology generated over 900 match alerts for a single [enrolled image of person of interest]''} ~\cite{ftc_v_riteaid_2023_complaint}. The FTC filed a complaint and a stipulated order that proposed comprehensive safeguards, including banning Rite Aid from using facial recognition AI for five years~\cite{ftc_v_riteaid_2024_order}.

Academic research has long known about the societal impact of biased scores, such as the use of credit scores for hiring and insurance~\cite{o2017weapons}. Research to date has focused on methods to fix the biased score. For example, if credit scores are known to be biased against Black applicants, various methods exist to address this, such as decoupled classifiers~\cite{dwork2018decoupled} or representation learning~\cite{zemel2013learning}.  Other work has already studied how interventions to combat algorithmic bias can have cascading effects~\cite{ghai2022cascaded}. However, there is an urgent need to study the statistical consequences of the integration of biased scores and whether known biases can be ameliorated. In its current state, the lack of research frameworks and measurement techniques for these cascading effects leaves enforcement agencies ill-equipped to investigate such compounded discrimination, even as the practice becomes increasingly common in commercial applications.

To address cascading algorithmic bias, ML researchers should focus on three critical areas: (1) developing formal frameworks to model and quantify how discrimination compounds across sequential algorithmic systems, accounting for both direct effects and indirect interactions through proxy variables; (2) creating practical tools that can trace the propagation of bias through complex scoring networks while maintaining computational feasibility for real-world applications; and (3) designing intervention techniques that can detect and mitigate discriminatory effects at both individual decision points and across entire chains of algorithmic systems. These advances would enable regulators and investigators to better understand, document, and address compounded discrimination in increasingly interconnected automated decision systems.


\subsection{Quantifying Disparate Impact}
To demonstrate disparate impact, investigators must provide statistical evidence showing that a practice disproportionately affects protected groups. Courts typically rely on two key metrics: the four-fifths rule (as discussed earlier) and statistical significance tests showing that disparities are unlikely to occur by chance. This evidence often combines direct outcome comparisons (e.g., loan approval rates by race) with regression analyses that control for legitimate business factors. For algorithmic systems, plaintiffs employ additional methods, including audit studies that compare outcomes across identical profiles varying only protected characteristics, analysis of proxy variables that may correlate with protected status, and evaluation of model performance across demographic subgroups. Courts generally require both statistical significance (e.g., 95\% confidence intervals) and practical significance, along with evidence that observed disparities are causally linked to the challenged practice rather than external factors. This can be a challenging requirement to meet in settings with complex algorithmic systems. 

In 2021, the New York State Department of Financial Services (NYDFS) investigated Apple based on consumer complaints about the Apple Card and the potential for algorithmic discrimination. After a lengthy investigation, the report found no evidence of unlawful discrimination against applicants under fair lending law. Part of the investigation focused on individuals who sent discrimination complaints to the NYDFS: \textit{``In each instance, the Bank was able to identify the factors that led to the credit decisions, such as credit score, indebtedness, income, credit utilization, missed payments, and other credit history elements''}~\cite{applecardinvestigation}. Using regression analysis, investigators were able to determine that the Apple Card lending policy---and the underlying statistical model---would not produce disparate impacts~\cite{applecardinvestigation}.

While regression analysis proved sufficient for evaluating the Apple Card's relatively straightforward lending decisions, more sophisticated approaches are often needed for complex algorithmic systems. As an example, in 2022, the Department of Justice alleged that Meta's housing ad delivery algorithms used characteristics protected under the Fair Housing Act. The algorithmic complexity of ad targeting required a more advanced settlement approach than traditional disparate impact remedies. As a result, the settlement agreement describes, \textit{``Meta will develop a system to reduce variances in Ad Impressions between Eligible Audiences and Actual Audiences, which the United States alleges are introduced by Meta's ad delivery system, for sex and estimated race/ethnicity''}~\cite{meta2024}. This ``variance reduction system'' seeks to address demographic skew in ad delivery---a novel technical approach necessitated by the limitations of traditional impact analysis. 

Our analysis suggests several critical areas for future research in quantifying algorithmic disparate impact. As algorithmic systems become more complex, they require sophisticated measurement approaches that can: (1) isolate interaction effects between protected attributes and other variables, (2) track temporal drift in model behavior and fairness metrics, and (3) model complex downstream economic consequences. Machine learning concepts such as distribution shift, causal inference, and feature attribution are crucial for these research questions. However, more work is needed to develop practical solutions. For example, in lending discrimination, impact quantification must account for both immediate effects of loan denials and long-term consequences for wealth accumulation and inter-generational mobility. Methods from econometrics and causal inference may help address this change. While prior works advocate for causal frameworks in measuring disparate impact~\cite{xiang2019legal,xiang2020reconciling}, the field lacks validated methodologies and case studies demonstrating how to compute these measures in ways that both withstand legal scrutiny and support specific damage calculations. The few examples that do exist rely on quasi-experimental methods that are often infeasible to implement with the available data \cite{arnold2021measuring}. Recent work has highlighted that regulatory approaches must consider threshold variations across different populations~\cite{meursault2024threshold} and has begun developing frameworks for explainable fairness in regulatory auditing contexts~\cite{oneil2024explainable}.
This methodological gap significantly hampers enforcement actions seeking concrete remedies for affected populations.

\subsection{Information Barriers in Algorithm Investigations}
\label{sec:information-barriers}
Civil enforcement agencies often work with incomplete and potentially obscured information about deployed systems when investigating algorithmic discrimination.
While enforcement agencies can compel data disclosure through administrative subpoenas, civil investigative demands, or discovery requests in litigation, companies often provide incomplete or inconsistent information due to poor documentation or strategic withholding. Sometimes, even the existence of algorithmic decision-making is undisclosed from affected individuals. This information asymmetry creates a formidable barrier to investigation: individuals cannot challenge discriminatory practices they do not know exist. 


Other sectors demonstrate how data transparency requirements can enable effective algorithmic investigations. The Home Mortgage Disclosure Act (HMDA) mandates mortgage lenders to report their data, which proved crucial in an ongoing class action lawsuit against Navy Federal Credit Union. When mortgage applicants filed a 2023 class action alleging systematic racial discrimination in the credit union's underwriting policies, they cited evidence from earlier National Credit Union Administration (NCUA) research: \textit{``In November 2022, the [NCUA] published
research using publicly-available 2020 and 2021 HMDA data to assess racial and ethnic disparities in mortgage
lending by credit unions''}~\cite{oliver2024navy}. This illustrates how legal requirements for data sharing can facilitate the investigation of algorithmic discrimination.

\begin{figure*}[t]
    \centering
    \includegraphics[width=0.8\textwidth]{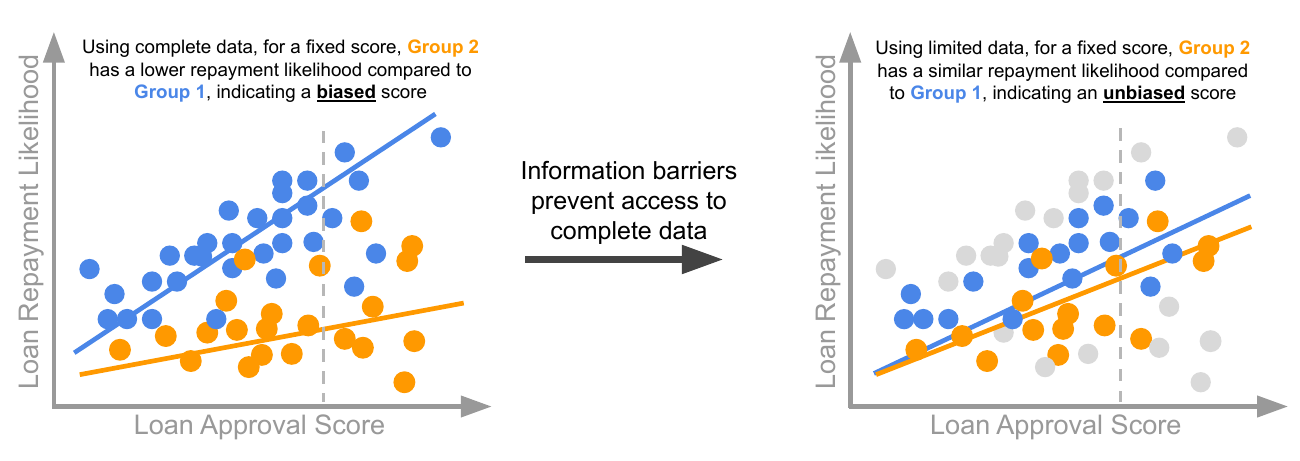}
    \caption{Illustration of how information barriers can affect algorithmic investigations of discrimination. Information barriers may include only being able to access data from certain years due to improper data retention protocols by the company under investigation.}
    \label{fig:information-barriers}
\end{figure*}

Information barriers create unique technical challenges distinct from academic fairness research settings, where current fairness metrics and auditing approaches typically assume access to complete training data. These methods may prove insufficient in adversarial conditions. For a simplified example, Figure~\ref{fig:information-barriers} demonstrates how the findings of an algorithmic investigation may change depending on which cohorts of training and outcome data are available. To address the challenge of information barriers, investigators must develop robust methods to: (1) reconstruct historical model versions from code fragments and deployment logs, (2) infer training data distributions from partial samples, and (3) validate findings using only available system inputs and outputs. The field lacks systematic methodologies for conducting rigorous bias investigations under information constraints, and this gap significantly impacts enforcement effectiveness. Research on adversarial algorithmic bias auditing could help determine the minimal set of records and information needed for an algorithmic audit. 


\subsection{Missing Protected Group Information}

Without accurate demographic information, investigators may struggle to establish statistical evidence of discriminatory patterns and face heightened burdens of proof in legal proceedings. The lack of reliable race and ethnicity data particularly hampers class certification efforts, as courts require clear methods for identifying affected individuals. 
As an extreme case, certain sectors may forbid the consideration---and therefore collection---of these groups, such as the Equal Credit Reporting Act, which prohibits creditors from considering information about protected attributes in any aspect of a credit transaction~\cite{regB2009,reg_b_2021}. The lack of recorded information affected the Allstate 2007 settlement regarding allegedly racially discriminatory insurance premiums because \textit{``Allstate does not maintain information on the race of its policyholders,''} and therefore notice could not be given to prospective class members~\cite{dehoyos2007}.

To address these data limitations, investigators have leveraged methods to inpute protected group status based on available data, commonly using name and location information. As one example, Bayesian Improved Surname Geocoding (BISG) is a statistical method that uses census data,  surname, and geographic location data to probabilistically estimate an individual's race and ethnicity~\cite{adjaye2014using}. In the settlement agreement between the US and Meta regarding housing algorithmic discrimination, BISG is directly mentioned as a method to address missing protected attribute data for auditing housing ad-targeting algorithms: \textit{``For estimated race/ethnicity, measurements will be based on information estimating race/ethnicity using a privacy-enhanced version of [BISG].''}~\cite{meta2024}. 
While methods exist for demographic imputation, these introduce complex statistical uncertainties that complicate enforcement actions. Investigators face several technical challenges, including propagating group imputation uncertainty through disparate impact calculations and establishing confidence bounds on fairness metrics under partial identification. 

In many ways, the lack of protected group information and the statistical methods to impute it are closely related to research gaps previously discussed in this work. As in the case of cascading algorithmic bias, investigators must grapple with how flawed data and algorithms may magnify biases. In this case, however, auditors can choose which methods to use for imputing group information—or whether to use another method to determine disparate impact altogether. Like when information barriers are present, we again have incomplete data that may stymie analysis. The difference in the case of missing protected group information is that discriminatory analysis rests on protected group information, meaning investigators can ascertain immediately when the information is missing and address the data gap directly. Due to the interconnected nature of the methodological components of algorithmic investigations, addressing any of these research gaps is likely to benefit other gaps as well.


Recent work has begun addressing these challenges, such as sensitivity analysis frameworks for methods like BISG, which predict protected class from proxy variables like surname and location using an auxiliary dataset, e.g., census data~\cite{kallus2022assessing}. However, significant gaps remain in handling heterogeneous data quality across subgroups and accounting for spatial/temporal variation in proxy accuracy. While some companies justify data gaps using ``fairness through blindness'' principles, this primarily obscures discrimination by making violations harder to detect and prove. This creates pressing needs for methodologies that can: (1) leverage multiple imperfect proxy sources through ensemble methods, (2) provide formal guarantees on bias detection power under specified missing data mechanisms, and (3) quantify minimum detectability thresholds for different types of algorithmic discrimination given available data quality.

\section{Discussion}
This work analyzes civil enforcement actions against algorithmic discrimination, surfacing five critical research gaps: cascading bias, finding less discriminatory alternatives, quantifying disparate impact, incomplete documentation, and protected group imputation. These gaps highlight a disconnect between academic fairness research and enforcement needs, presenting concrete opportunities for developing tools to strengthen legal actions. Particularly urgent are methods for reconstructing algorithmic behavior from partial information, quantifying discrimination under uncertainty, and efficiently exploring less biased alternatives.

Our analysis has many limitations. Our focus on civil enforcement excludes legislative proposals and protections that have not resulted in enforcement action, though this scope maintains practical applicability. While we primarily considered claims involving private sector algorithms, government AI systems warrant similar scrutiny, as evidenced by controversial facial recognition deployments in law enforcement~\cite{engstrom2020government, bbcClearviewUsed}. Additionally, our reliance on public documents about confidential investigations was, for the reasons  described above, necessarily not exhaustive and did not capture all enforcement challenges.

Looking forward, the emergence of large language models and generative AI introduces new complexities for fairness enforcement, from opaque training processes and dynamic outputs to challenges in defining and measuring bias in generative systems~\cite{pfohl2024toolbox,gallegos2024bias}. We advocate for deeper collaboration between ML researchers and enforcement agencies through joint case studies of algorithmic discrimination investigations. Such partnerships could validate new fairness methodologies under real-world constraints while equipping agencies with technical capabilities to strengthen enforcement. As AI systems grow more sophisticated, bridging research and enforcement communities will become increasingly critical for ensuring algorithmic fairness.

\section*{Acknowledgements}
The authors thank Kristie Chamorro, Skylar Cushing, and Juliette Draper for help with the case and literature reviews. The authors thank Deb Raji and Ziad Obermeyer for feedback and advice, and Iris Cisneros for copyediting assistance.

\bibliography{BIB}

\begin{thebibliography}{99}
\providecommand{\natexlab}[1]{#1}

\bibitem[{Adjaye-Gbewonyo et~al.(2014)Adjaye-Gbewonyo, Bednarczyk, Davis, and Omer}]{adjaye2014using}
Adjaye-Gbewonyo, D.; Bednarczyk, R.~A.; Davis, R.~L.; and Omer, S.~B. 2014.
\newblock Using the Bayesian Improved Surname Geocoding Method (BISG) to create a working classification of race and ethnicity in a diverse managed care population: a validation study.
\newblock \emph{Health services research}, 49(1): 268--283.

\bibitem[{{American Civil Liberties Union Foundation}(2024)}]{aclu2024aon}
{American Civil Liberties Union Foundation}. 2024.
\newblock {In the Matter of Aon Consulting, Inc.: Complaint and Request for Investigation, Injunction, and Other Relief}.
\newblock Filed with the Federal Trade Commission, Washington, DC.

\bibitem[{Arnold, Dobbie, and Hull(2021)}]{arnold2021measuring}
Arnold, D.; Dobbie, W.; and Hull, P. 2021.
\newblock Measuring racial discrimination in algorithms.
\newblock In \emph{AEA Papers and Proceedings}, volume 111, 49--54.

\bibitem[{Arya et~al.(2021)Arya, Bellamy, Chen, Dhurandhar, Hind, Hoffman, Houde, Liao, Luss, Mojsilovi{\'c} et~al.}]{arya2021ai}
Arya, V.; Bellamy, R.~K.; Chen, P.-Y.; Dhurandhar, A.; Hind, M.; Hoffman, S.~C.; Houde, S.; Liao, Q.~V.; Luss, R.; Mojsilovi{\'c}, A.; et~al. 2021.
\newblock Ai explainability 360 toolkit.
\newblock In \emph{Proceedings of the 3rd ACM India Joint International Conference on Data Science \& Management of Data (8th ACM IKDD CODS \& 26th COMAD)}, 376--379.

\bibitem[{Balkin and Siegel(2003)}]{balkin2003american}
Balkin, J.~M.; and Siegel, R.~B. 2003.
\newblock The American civil rights tradition: Anticlassification or antisubordination.
\newblock \emph{Issues in Legal Scholarship}, 2(11).

\bibitem[{Barocas, Hardt, and Narayanan(2023)}]{barocas2023fairness}
Barocas, S.; Hardt, M.; and Narayanan, A. 2023.
\newblock \emph{Fairness and machine learning: Limitations and opportunities}.
\newblock MIT press.

\bibitem[{Barocas and Selbst(2016)}]{barocas2016big}
Barocas, S.; and Selbst, A.~D. 2016.
\newblock Big data's disparate impact.
\newblock \emph{Cal. L. Rev.}, 104: 671.

\bibitem[{Bartlett et~al.(2021)Bartlett, Morse, Wallace, and Stanton}]{bartlett2021algorithmic}
Bartlett, R.; Morse, A.; Wallace, N.; and Stanton, R. 2021.
\newblock Algorithmic Discrimination and Input Accountability under the Civil Rights Act.
\newblock \emph{Berkeley Tech. L. J.}, 36: 675.

\bibitem[{Bedayn(2024)}]{apnewsAttemptsRegulate}
Bedayn, J. 2024.
\newblock {A}ttempts to regulate {A}{I}’s hidden hand in {A}mericans’ lives flounder in {U}{S} statehouses.
\newblock \url{https://apnews.com/article/artificial-intelligence-bias-discrimination-regulation-ai-ff1d0860663723079aac3666b38f2320}.
\newblock [Accessed 19-01-2025].

\bibitem[{Bent(2019)}]{bent2019algorithmic}
Bent, J.~R. 2019.
\newblock Is algorithmic affirmative action legal.
\newblock \emph{Geo. LJ}, 108: 803.

\bibitem[{Black et~al.(2024)Black, Koepke, Kim, Barocas, and Hsu}]{black2024legal}
Black, E.; Koepke, J.~L.; Kim, P.; Barocas, S.; and Hsu, M. 2024.
\newblock Less Discriminatory Algorithms.
\newblock \emph{Geo. L. J.}, 113: 53.

\bibitem[{Breiman(2001)}]{breiman2001statistical}
Breiman, L. 2001.
\newblock Statistical modeling: The two cultures (with comments and a rejoinder by the author).
\newblock \emph{Statistical science}, 16(3): 199--231.

\bibitem[{Buolamwini and Gebru(2018)}]{buolamwini2018gender}
Buolamwini, J.; and Gebru, T. 2018.
\newblock Gender shades: Intersectional accuracy disparities in commercial gender classification.
\newblock In \emph{Conference on fairness, accountability and transparency}, 77--91. PMLR.

\bibitem[{Calmon et~al.(2017)Calmon, Wei, Vinzamuri, Ramamurthy, and Varshney}]{calmon2017optimized}
Calmon, F.; Wei, D.; Vinzamuri, B.; Ramamurthy, K.~N.; and Varshney, K.~R. 2017.
\newblock Optimized Pre-Processing for Discrimination Prevention.
\newblock In \emph{Advances in Neural Information Processing Systems}, 3995--4004.

\bibitem[{CFPB(2011)}]{regB2009}
CFPB. 2011.
\newblock Equal Credit Opportunity Act (Regulation B).
\newblock Rules concerning evaluation of applications.

\bibitem[{CFPB(2023)}]{consumerfinanceCFPBIssues}
CFPB. 2023.
\newblock {C}{F}{P}{B} {I}ssues {G}uidance on {C}redit {D}enials by {L}enders {U}sing {A}rtificial {I}ntelligence.
\newblock \url{https://www.consumerfinance.gov/about-us/newsroom/cfpb-issues-guidance-on-credit-denials-by-lenders-using-artificial-intelligence/}.
\newblock [Accessed 20-01-2025].

\bibitem[{Chen, Johansson, and Sontag(2018)}]{chen2018my}
Chen, I.~Y.; Johansson, F.~D.; and Sontag, D. 2018.
\newblock Why Is My Classifier Discriminatory?
\newblock In \emph{Neural Information Processing Systems (NeurIPS) 2018}, 3543--3554.

\bibitem[{Chen et~al.(2020)Chen, Pierson, Rose, Joshi, Ferryman, and Ghassemi}]{chen2020ethical}
Chen, I.~Y.; Pierson, E.; Rose, S.; Joshi, S.; Ferryman, K.; and Ghassemi, M. 2020.
\newblock Ethical Machine Learning in Health.
\newblock \emph{Annual Review of Biomedical Data Science 4}.

\bibitem[{Chen, Szolovits, and Ghassemi(2019)}]{chen2019can}
Chen, I.~Y.; Szolovits, P.; and Ghassemi, M. 2019.
\newblock Can AI Help Reduce Disparities in General Medical and Mental Health Care?
\newblock \emph{AMA Journal of Ethics}, 21(2): 167--179.

\bibitem[{Chien(2020)}]{umichquotAmericaapossPaper}
Chien, C. 2020.
\newblock {A}merica's {P}aper {P}risons: {T}he {S}econd {C}hance {G}ap.
\newblock \emph{Mich. L. Rev.}, 119: 519.

\bibitem[{Chouldechova(2017)}]{chouldechova2017fair}
Chouldechova, A. 2017.
\newblock Fair prediction with disparate impact: A study of bias in recidivism prediction instruments.
\newblock \emph{Big data}, 5(2): 153--163.

\bibitem[{Clayton and Derico(2023)}]{bbcClearviewUsed}
Clayton, J.; and Derico, B. 2023.
\newblock {C}learview {A}{I} used nearly 1m times by {U}{S} police, it tells the {B}{B}{C} --- bbc.com.
\newblock \url{https://www.bbc.com/news/technology-65057011}.
\newblock [Accessed 20-01-2025].

\bibitem[{{Co. General Assembly}(2024)}]{coloradoConsumerProtections}
{Co. General Assembly}. 2024.
\newblock {Consumer Protections for Artificial Intelligence}.

\bibitem[{{Consumer Financial Protection Bureau}(2018)}]{reg_b_2021}
{Consumer Financial Protection Bureau}. 2018.
\newblock {Rules Concerning Evaluation of Applications}.
\newblock 12 C.F.R. \S~1002.6(b)(9).

\bibitem[{Corbett-Davies et~al.(2023)Corbett-Davies, Gaebler, Nilforoshan, Shroff, and Goel}]{corbett2023measure}
Corbett-Davies, S.; Gaebler, J.~D.; Nilforoshan, H.; Shroff, R.; and Goel, S. 2023.
\newblock The measure and mismeasure of fairness.
\newblock \emph{The Journal of Machine Learning Research}, 24(1): 14730--14846.

\bibitem[{Dastin(2018)}]{dastin2018}
Dastin, J. 2018.
\newblock Insight - Amazon scraps secret AI recruiting tool that showed bias against women.
\newblock \url{https://www.reuters.com/article/world/insight-amazon-scraps-secret-ai-recruiting-tool-that-showed-bias-against-women-idUSKCN1MK0AG/}.
\newblock [Accessed 18-01-2025].

\bibitem[{DeHoyos(2007)}]{dehoyos2007}
DeHoyos. 2007.
\newblock Settlement Agreement in DeHoyos v. {A}llstate Corp.
\newblock \emph{F.R.D.}, 240: 269.
\newblock No. CIV.A. SA01CA1010FB, Decided Feb 21, 2007.

\bibitem[{{DeHoyos et al.}(2001)}]{dehoyos2001}
{DeHoyos et al.} 2001.
\newblock {Complaint in DeHoyos v. Allstate Insurance Company}.
\newblock Civil Action No. SA-01-CA-1010-FB, United States District Court, Western District of Texas, San Antonio Division.

\bibitem[{{DOJ}(2024)}]{meta2024}
{DOJ}. 2024.
\newblock Settlement Agreement, United {S}tates of {A}merica {DOJ}, v. {M}eta {P}latforms, Inc. f/k/a {F}acebook, Inc.,.
\newblock Case No. 1:22-cv-05187, United States District Court Southern District of New York.

\bibitem[{Dwork et~al.(2012)Dwork, Hardt, Pitassi, Reingold, and Zemel}]{dwork2012fairness}
Dwork, C.; Hardt, M.; Pitassi, T.; Reingold, O.; and Zemel, R. 2012.
\newblock Fairness through awareness.
\newblock In \emph{Proceedings of the 3rd Innovations in Theoretical Computer Science Conference}, 214--226. ACM.

\bibitem[{Dwork et~al.(2018)Dwork, Immorlica, Kalai, and Leiserson}]{dwork2018decoupled}
Dwork, C.; Immorlica, N.; Kalai, A.~T.; and Leiserson, M. 2018.
\newblock Decoupled classifiers for group-fair and efficient machine learning.
\newblock In \emph{Conference on fairness, accountability and transparency}, 119--133. PMLR.

\bibitem[{{EEOC}(2023)}]{eeocSelectIssues}
{EEOC}. 2023.
\newblock {Select Issues: Assessing Adverse Impact in Software, Algorithms, and Artificial Intelligence Used in Employment Selection Procedures Under Title VII of the Civil Rights Act of 1964}.

\bibitem[{{Electronic Privacy Information Center}(2019)}]{epic2019hirevue}
{Electronic Privacy Information Center}. 2019.
\newblock {In the Matter of HireVue, Inc.: Complaint and Request for Investigation, Injunction, and Other Relief}.
\newblock Filed with the Federal Trade Commission, Washington, DC.

\bibitem[{Engstrom et~al.(2020)Engstrom, Ho, Sharkey, and Cu{\'e}llar}]{engstrom2020government}
Engstrom, D.~F.; Ho, D.~E.; Sharkey, C.~M.; and Cu{\'e}llar, M.-F. 2020.
\newblock Government by algorithm: Artificial intelligence in federal administrative agencies.
\newblock \emph{NYU School of Law, Public Law Research Paper}, (20-54).

\bibitem[{{Estate of Gene B. Lokken et al.}(2024)}]{lokken2024}
{Estate of Gene B. Lokken et al.} 2024.
\newblock First Amended Class Action Complaint, Lokken v. UnitedHealth Group, Inc,.
\newblock {United States District Court for the District of Minnesota, Case No. 23-cv-3514}.

\bibitem[{EU(2024)}]{artificialintelligenceactArtificialIntelligence}
EU. 2024.
\newblock {E}{U} {A}rtificial {I}ntelligence {A}ct.
\newblock \url{https://artificialintelligenceact.eu/}.
\newblock [Accessed 20-01-2025].

\bibitem[{Feldman et~al.(2015)Feldman, Friedler, Moeller, Scheidegger, and Venkatasubramanian}]{feldman2015certifying}
Feldman, M.; Friedler, S.~A.; Moeller, J.; Scheidegger, C.; and Venkatasubramanian, S. 2015.
\newblock Certifying and removing disparate impact.
\newblock In \emph{Proceedings of the 21th ACM SIGKDD International Conference on Knowledge Discovery and Data Mining}, 259--268. ACM.

\bibitem[{Fisher, Rudin, and Dominici(2019)}]{fisher2019all}
Fisher, A.; Rudin, C.; and Dominici, F. 2019.
\newblock All models are wrong, but many are useful: Learning a variable's importance by studying an entire class of prediction models simultaneously.
\newblock \emph{Journal of Machine Learning Research}, 20(177): 1--81.

\bibitem[{Friedler, Scheidegger, and Venkatasubramanian(2021)}]{friedler2021possibility}
Friedler, S.~A.; Scheidegger, C.; and Venkatasubramanian, S. 2021.
\newblock The (im) possibility of fairness: Different value systems require different mechanisms for fair decision making.
\newblock \emph{Communications of the ACM}, 64(4): 136--143.

\bibitem[{FTC(2023)}]{ftc_v_riteaid_2023_complaint}
FTC. 2023.
\newblock Complaint for Permanent Injunction and Other Relief, Federal Trade Commission v. Rite Aid Corporation and Rite Aid Hdqtrs. Corp.
\newblock United States District Court for the Eastern District of Pennsylvania, Case No. 2:23-cv-5023.

\bibitem[{{FTC}(2024)}]{ftc_v_riteaid_2024_order}
{FTC}. 2024.
\newblock {Stipulated Order for Permanent Injunction and Other Relief, Federal Trade Commission v. Rite Aid Corporation and Rite Aid Hdqtrs. Corp.}
\newblock United States District Court for the Eastern District of Pennsylvania, Case No. 2:23-cv-5023.

\bibitem[{Gallegos et~al.(2024)Gallegos, Rossi, Barrow, Tanjim, Kim, Dernoncourt, Yu, Zhang, and Ahmed}]{gallegos2024bias}
Gallegos, I.~O.; Rossi, R.~A.; Barrow, J.; Tanjim, M.~M.; Kim, S.; Dernoncourt, F.; Yu, T.; Zhang, R.; and Ahmed, N.~K. 2024.
\newblock Bias and fairness in large language models: A survey.
\newblock \emph{Computational Linguistics}, 1--79.

\bibitem[{Ghai, Mishra, and Mueller(2022)}]{ghai2022cascaded}
Ghai, B.; Mishra, M.; and Mueller, K. 2022.
\newblock Cascaded debiasing: Studying the cumulative effect of multiple fairness-enhancing interventions.
\newblock In \emph{Proceedings of the 31st ACM International Conference on Information \& Knowledge Management}, 3082--3091.

\bibitem[{Gillis, Meursault, and Ustun(2024)}]{gillis2024operationalizing}
Gillis, T.~B.; Meursault, V.; and Ustun, B. 2024.
\newblock Operationalizing the Search for Less Discriminatory Alternatives in Fair Lending.
\newblock In \emph{Proceedings of the 2024 ACM Conference on Fairness, Accountability, and Transparency}, FAccT '24, 377--387. Rio de Janeiro, Brazil: ACM.

\bibitem[{Green(2022)}]{green2022flaws}
Green, B. 2022.
\newblock The flaws of policies requiring human oversight of government algorithms.
\newblock \emph{Computer Law \& Security Review}, 45: 105681.

\bibitem[{{Griggs}(1971)}]{griggs1971}
{Griggs}. 1971.
\newblock {Griggs v. Duke Power Co.}
\newblock 401 U.S. 424, Supreme Court of the United States.

\bibitem[{Groves et~al.(2024)Groves, Metcalf, Kennedy, Vecchione, and Strait}]{groves2024auditing}
Groves, L.; Metcalf, J.; Kennedy, A.; Vecchione, B.; and Strait, A. 2024.
\newblock Auditing work: Exploring the New York City algorithmic bias audit regime.
\newblock In \emph{The 2024 ACM Conference on Fairness, Accountability, and Transparency}, 1107--1120.

\bibitem[{Hajian and Domingo-Ferrer(2013)}]{hajian2013methodology}
Hajian, S.; and Domingo-Ferrer, J. 2013.
\newblock A methodology for direct and indirect discrimination prevention in data mining.
\newblock \emph{IEEE transactions on knowledge and data engineering}, 25(7): 1445--1459.

\bibitem[{Hardt, Price, and Srebro(2016)}]{hardt2016equality}
Hardt, M.; Price, E.; and Srebro, N. 2016.
\newblock Equality of opportunity in supervised learning.
\newblock \emph{Advances in neural information processing systems}, 29.

\bibitem[{Heidari et~al.(2018)Heidari, Ferrari, Gummadi, and Krause}]{heidari2018fairness}
Heidari, H.; Ferrari, C.; Gummadi, K.; and Krause, A. 2018.
\newblock Fairness behind a veil of ignorance: A welfare analysis for automated decision making.
\newblock \emph{Advances in neural information processing systems}, 31.

\bibitem[{Hellman(2020)}]{hellman2020measuring}
Hellman, D. 2020.
\newblock Measuring algorithmic fairness.
\newblock \emph{Va. L. Rev.}, 106: 811.

\bibitem[{Henderson et~al.(2024)Henderson, Hu, Diab, and Pineau}]{henderson2024rethinking}
Henderson, P.; Hu, J.; Diab, M.; and Pineau, J. 2024.
\newblock Rethinking Machine Learning Benchmarks in the Context of Professional Codes of Conduct.
\newblock In \emph{Proceedings of the Symposium on Computer Science and Law}, 109--120.

\bibitem[{Ho and Xiang(2020)}]{ho2020affirmative}
Ho, D.~E.; and Xiang, A. 2020.
\newblock Affirmative algorithms: The legal grounds for fairness as awareness.
\newblock \emph{U. Chi. L. Rev. Online}, 134.

\bibitem[{Huskey(2022)}]{huskey2022complaint}
Huskey, J. 2022.
\newblock {Class Action Complaint, Huskey v. State Farm Fire \& Casualty Company}.
\newblock U.S. District Court for the Northern District of Illinois, Eastern Division, No. 22-cv-7014.

\bibitem[{{Ill. General Assembly}(2020)}]{il-ai-video-2020}
{Ill. General Assembly}. 2020.
\newblock Artificial {I}ntelligence {V}ideo {I}nterview {A}ct.

\bibitem[{{Ill. General Assembly}(2023)}]{ilgaIllinoisGeneral}
{Ill. General Assembly}. 2023.
\newblock {I}llinois {H}ouse {B}ill 3773.

\bibitem[{Jacobs and Sparrow(2000)}]{jacobs2000fair}
Jacobs, R.~C.; and Sparrow, R.~S. 2000.
\newblock Fair Housing Act Guidance Memorandum of Understanding of the Treasury, HUD, and Justice Departments.
\newblock \emph{J. Affordable Hous. \& Cmty. Dev. L.}, 10: 16.

\bibitem[{Kallus, Mao, and Zhou(2022)}]{kallus2022assessing}
Kallus, N.; Mao, X.; and Zhou, A. 2022.
\newblock Assessing algorithmic fairness with unobserved protected class using data combination.
\newblock \emph{Management Science}, 68(3): 1959--1981.

\bibitem[{Kamiran, Calders, and Pechenizkiy(2010)}]{kamiran2010discrimination}
Kamiran, F.; Calders, T.; and Pechenizkiy, M. 2010.
\newblock Discrimination aware decision tree learning.
\newblock In \emph{Data Mining (ICDM), 2010 IEEE 10th International Conference on}, 869--874. IEEE.

\bibitem[{Kamishima, Akaho, and Sakuma(2011)}]{kamishima2011fairness}
Kamishima, T.; Akaho, S.; and Sakuma, J. 2011.
\newblock Fairness-aware learning through regularization approach.
\newblock In \emph{Data Mining Workshops (ICDMW), 2011 IEEE 11th International Conference on}, 643--650. IEEE.

\bibitem[{Kim(2022)}]{kim2022race}
Kim, P.~T. 2022.
\newblock Race-aware algorithms: Fairness, nondiscrimination and affirmative action.
\newblock \emph{Cal. L. Rev.}, 110: 1539.

\bibitem[{Kleinberg, Mullainathan, and Raghavan(2016)}]{kleinberg2016inherent}
Kleinberg, J.; Mullainathan, S.; and Raghavan, M. 2016.
\newblock Inherent trade-offs in the fair determination of risk scores.
\newblock \emph{arXiv preprint arXiv:1609.05807}.

\bibitem[{Kroll et~al.(2017)Kroll, Huey, Barocas, Felten, Reidenberg, Robinson, and Yu}]{kroll165accountable}
Kroll, J.~A.; Huey, J.; Barocas, S.; Felten, E.~W.; Reidenberg, J.~R.; Robinson, D.~G.; and Yu, H. 2017.
\newblock Accountable algorithms.
\newblock \emph{U. Penn. L. Rev.}, 165: 633.

\bibitem[{Kusner et~al.(2017)Kusner, Loftus, Russell, and Silva}]{kusner2017counterfactual}
Kusner, M.~J.; Loftus, J.; Russell, C.; and Silva, R. 2017.
\newblock Counterfactual fairness.
\newblock In \emph{Advances in Neural Information Processing Systems}, 4069--4079.

\bibitem[{Lipton, McAuley, and Chouldechova(2018)}]{lipton2018does}
Lipton, Z.; McAuley, J.; and Chouldechova, A. 2018.
\newblock Does mitigating ML's impact disparity require treatment disparity?
\newblock \emph{Advances in neural information processing systems}, 31.

\bibitem[{Metcalf et~al.(2021)Metcalf, Moss, Watkins, Singh, and Elish}]{metcalf2021algorithmic}
Metcalf, J.; Moss, E.; Watkins, E.~A.; Singh, R.; and Elish, M.~C. 2021.
\newblock Algorithmic impact assessments and accountability: The co-construction of impacts.
\newblock In \emph{Proceedings of the 2021 ACM conference on fairness, accountability, and transparency}, 735--746.

\bibitem[{Meursault et~al.(2025)Meursault, Moulton, Santucci, and Schor}]{meursault2024threshold}
Meursault, V.; Moulton, D.; Santucci, L.; and Schor, N. 2025.
\newblock One threshold doesn't fit all: {Tailoring} machine learning predictions of consumer default for lower-income areas.
\newblock \emph{Journal of Policy Analysis and Management}, (3): 792--815.

\bibitem[{Milstein(2024)}]{cohenmilsteinAIRelatedDiscrimination}
Milstein, C. 2024.
\newblock {A}{I}-{R}elated {D}iscrimination {S}uit {R}eaches {F}inal {S}ettlement.
\newblock \url{https://www.cohenmilstein.com/class-action-lawsuit-on-ai-related-discrimination-reaches-final-settlement/}.
\newblock [Accessed 19-01-2025].

\bibitem[{Mobley(2016)}]{mobley2016}
Mobley. 2016.
\newblock Settlement in Mobley v. {Facebook}, {Inc.}
\newblock 5:16-cv-06440.

\bibitem[{{Mobley}(2024)}]{mobley2024workday}
{Mobley}. 2024.
\newblock Order Granting in Part and Denying In Part Motion to Dismiss, Mobley v. {Workday} {Inc.}
\newblock 740 F. Supp. 3d 796, Northern District of California.

\bibitem[{Mulligan et~al.(2019)Mulligan, Kroll, Kohli, and Wong}]{mulligan2019thing}
Mulligan, D.~K.; Kroll, J.~A.; Kohli, N.; and Wong, R.~Y. 2019.
\newblock This thing called fairness: Disciplinary confusion realizing a value in technology.
\newblock \emph{Proceedings of the ACM on Human-Computer Interaction}, 3(CSCW): 1--36.

\bibitem[{{National Fair Housing Alliance}(2018)}]{nfha_v_facebook_2018}
{National Fair Housing Alliance}. 2018.
\newblock First Amended Complaint, National Fair Housing Alliance v. Facebook, Inc.
\newblock United States District Court for the Southern District of New York, Case No. 1:18-cv-02689.

\bibitem[{Nelson(2010)}]{nelson2010credit}
Nelson, A.~A. 2010.
\newblock Credit scores, race, and residential sorting.
\newblock \emph{Journal of Policy Analysis and Management}, 29(1): 39--68.

\bibitem[{{New York State Department of Financial Services}(2021{\natexlab{a}})}]{applecardpressrelease}
{New York State Department of Financial Services}. 2021{\natexlab{a}}.
\newblock {DFS Issues Findings on the Apple Card and Its Underwriter Goldman Sachs Bank}.

\bibitem[{{New York State Department of Financial Services}(2021{\natexlab{b}})}]{applecardinvestigation}
{New York State Department of Financial Services}. 2021{\natexlab{b}}.
\newblock {Investigation of Apple Card and Goldman Sachs' Credit Card Practices}.

\bibitem[{Oliver(2024)}]{oliver2024navy}
Oliver. 2024.
\newblock Opinion, Oliver v. {N}avy {F}ederal {C}redit {U}nion.
\newblock United States District Court for the Eastern District of Virginia, Alexandria Division, Case No. 1:23-cv-1731, 2024 U.S. Dist. LEXIS 96704.

\bibitem[{O'Neil(2017)}]{o2017weapons}
O'Neil, C. 2017.
\newblock \emph{Weapons of math destruction: How big data increases inequality and threatens democracy}.
\newblock Crown.

\bibitem[{O'Neil, Sargeant, and Appel(2024)}]{oneil2024explainable}
O'Neil, C.; Sargeant, H.; and Appel, J. 2024.
\newblock Explainable Fairness in Regulatory Algorithmic Auditing.
\newblock \emph{W. Va. L. Rev.}, 127: 79.

\bibitem[{Pfohl et~al.(2024)Pfohl, Cole-Lewis, Sayres, Neal, Asiedu, Dieng, Tomasev, Rashid, Azizi, Rostamzadeh et~al.}]{pfohl2024toolbox}
Pfohl, S.~R.; Cole-Lewis, H.; Sayres, R.; Neal, D.; Asiedu, M.; Dieng, A.; Tomasev, N.; Rashid, Q.~M.; Azizi, S.; Rostamzadeh, N.; et~al. 2024.
\newblock A toolbox for surfacing health equity harms and biases in large language models.
\newblock \emph{Nature Medicine}, 1--11.

\bibitem[{Raji et~al.(2020)Raji, Smart, White, Mitchell, Gebru, Hutchinson, Smith-Loud, Theron, and Barnes}]{raji2020closing}
Raji, I.~D.; Smart, A.; White, R.~N.; Mitchell, M.; Gebru, T.; Hutchinson, B.; Smith-Loud, J.; Theron, D.; and Barnes, P. 2020.
\newblock Closing the AI accountability gap: Defining an end-to-end framework for internal algorithmic auditing.
\newblock In \emph{Proceedings of the 2020 conference on fairness, accountability, and transparency}, 33--44.

\bibitem[{Rodolfa et~al.(2020)Rodolfa, Salomon, Haynes, Mendieta, Larson, and Ghani}]{rodolfa2020case}
Rodolfa, K.~T.; Salomon, E.; Haynes, L.; Mendieta, I.~H.; Larson, J.; and Ghani, R. 2020.
\newblock Case study: predictive fairness to reduce misdemeanor recidivism through social service interventions.
\newblock In \emph{Proceedings of the 2020 Conference on Fairness, Accountability, and Transparency}, 142--153.

\bibitem[{Ross and Herman(2023)}]{statnewsUnitedHealthFaces}
Ross, C.; and Herman, B. 2023.
\newblock {U}nited{H}ealth faces class action lawsuit over algorithmic care denials in {M}edicare {A}dvantage plans --- statnews.com.
\newblock \url{https://www.statnews.com/2023/11/14/unitedhealth-class-action-lawsuit-algorithm-medicare-advantage/}.
\newblock [Accessed 18-01-2025].

\bibitem[{Roth(2024)}]{thevergeLandlordScreening}
Roth, E. 2024.
\newblock {A}{I} landlord screening tool will stop scoring low-income tenants after discrimination suit --- theverge.com.
\newblock \url{https://www.theverge.com/2024/11/20/24297692/ai-landlord-tool-saferent-low-income-tenants-discrimination-settlement}.
\newblock [Accessed 18-01-2025].

\bibitem[{Saleiro et~al.(2018)Saleiro, Kuester, Hinkson, London, Stevens, Anisfeld, Rodolfa, and Ghani}]{saleiro2018aequitas}
Saleiro, P.; Kuester, B.; Hinkson, L.; London, J.; Stevens, A.; Anisfeld, A.; Rodolfa, K.~T.; and Ghani, R. 2018.
\newblock Aequitas: A bias and fairness audit toolkit.
\newblock \emph{arXiv preprint arXiv:1811.05577}.

\bibitem[{Sandvig(2016)}]{sandvig_v_lynch_2016}
Sandvig, C. 2016.
\newblock {Complaint for Declaratory and Injunctive Relief, Sandvig v. Lynch}.
\newblock {United States District Court for the District of Columbia, Case No. 1:16-cv-01368}.

\bibitem[{Selbst et~al.(2019)Selbst, Boyd, Friedler, Venkatasubramanian, and Vertesi}]{selbst2019fairness}
Selbst, A.~D.; Boyd, D.; Friedler, S.~A.; Venkatasubramanian, S.; and Vertesi, J. 2019.
\newblock Fairness and abstraction in sociotechnical systems.
\newblock In \emph{Proceedings of the conference on fairness, accountability, and transparency}, 59--68.

\bibitem[{Seyyed-Kalantari et~al.(2021)Seyyed-Kalantari, Zhang, McDermott, Chen, and Ghassemi}]{seyyed2021underdiagnosis}
Seyyed-Kalantari, L.; Zhang, H.; McDermott, M.~B.; Chen, I.~Y.; and Ghassemi, M. 2021.
\newblock Underdiagnosis bias of artificial intelligence algorithms applied to chest radiographs in under-served patient populations.
\newblock \emph{Nature medicine}, 27(12): 2176--2182.

\bibitem[{Shen, Raji, and Chen(2024)}]{shen2024data}
Shen, J.~H.; Raji, I.~D.; and Chen, I.~Y. 2024.
\newblock The Data Addition Dilemma.
\newblock In \emph{Machine Learning for Healthcare Conference}. PMLR.

\bibitem[{United States District~Court(2023)}]{huskey2023}
United States District~Court, E.~D., N.D.~Illinois. 2023.
\newblock Huskey v. State Farm Fire \& Casualty Company.
\newblock No. 22 C 7014, 2023 WL 5848164 (N.D. Ill. Sep. 11, 2023).

\bibitem[{Wachter, Mittelstadt, and Russell(2020)}]{wachter2020bias}
Wachter, S.; Mittelstadt, B.; and Russell, C. 2020.
\newblock Bias preservation in machine learning: the legality of fairness metrics under EU non-discrimination law.
\newblock \emph{W. Va. L. Rev.}, 123: 735.

\bibitem[{Wang et~al.(2024)Wang, Wu, Ji, and Fu}]{wang2024algorithmic}
Wang, X.; Wu, Y.~C.; Ji, X.; and Fu, H. 2024.
\newblock Algorithmic Discrimination: Examining Its Types and Regulatory Measures with Emphasis on U.S. Legal Practices.
\newblock \emph{Frontiers in Artificial Intelligence, Security, Technology and Law}, 7.

\bibitem[{Watkins and Chen(2024)}]{watkins2024four}
Watkins, E.~A.; and Chen, J. 2024.
\newblock The four-fifths rule is not disparate impact: a woeful tale of epistemic trespassing in algorithmic fairness.
\newblock In \emph{The 2024 ACM Conference on Fairness, Accountability, and Transparency}, 764--775.

\bibitem[{{White House}(2025)}]{TrumpEO}
{White House}. 2025.
\newblock {US} {AI} {A}ction {P}lan.
\newblock \url{https://www.whitehouse.gov/wp-content/uploads/2025/07/Americas-AI-Action-Plan.pdf}.
\newblock [Accessed 25-07-2025].

\bibitem[{Woodworth et~al.(2017)Woodworth, Gunasekar, Ohannessian, and Srebro}]{woodworth2017learning}
Woodworth, B.; Gunasekar, S.; Ohannessian, M.~I.; and Srebro, N. 2017.
\newblock Learning Non-Discriminatory Predictors.
\newblock \emph{Conference On Learning Theory}.

\bibitem[{Wright et~al.(2024)Wright, Muenster, Vecchione, Qu, Cai, Smith, {Comm 2450 Student Investigators}, Metcalf, Matias et~al.}]{wright2024null}
Wright, L.; Muenster, R.~M.; Vecchione, B.; Qu, T.; Cai, P.; Smith, A.; {Comm 2450 Student Investigators}; Metcalf, J.; Matias, J.~N.; et~al. 2024.
\newblock Null Compliance: NYC Local Law 144 and the challenges of algorithm accountability.
\newblock In \emph{The 2024 ACM Conference on Fairness, Accountability, and Transparency}, 1701--1713.

\bibitem[{Xiang(2020)}]{xiang2020reconciling}
Xiang, A. 2020.
\newblock Reconciling legal and technical approaches to algorithmic bias.
\newblock \emph{Tenn. L. Rev.}, 88: 649.

\bibitem[{Xiang and Raji(2019)}]{xiang2019legal}
Xiang, A.; and Raji, I.~D. 2019.
\newblock On the legal compatibility of fairness definitions.
\newblock \emph{arXiv preprint arXiv:1912.00761}.

\bibitem[{Zafar et~al.(2017)Zafar, Valera, Gomez~Rodriguez, and Gummadi}]{zafar2017fairness}
Zafar, M.~B.; Valera, I.; Gomez~Rodriguez, M.; and Gummadi, K.~P. 2017.
\newblock Fairness beyond disparate treatment \& disparate impact: Learning classification without disparate mistreatment.
\newblock In \emph{Proceedings of the 26th International Conference on World Wide Web}, 1171--1180. International World Wide Web Conferences Steering Committee.

\bibitem[{Zemel et~al.(2013)Zemel, Wu, Swersky, Pitassi, and Dwork}]{zemel2013learning}
Zemel, R.~S.; Wu, Y.; Swersky, K.; Pitassi, T.; and Dwork, C. 2013.
\newblock Learning Fair Representations.
\newblock \emph{ICML (3)}, 28: 325--333.

\end{thebibliography}

\appendix

\section{Extended Table of Investigations}
\label{sec:appendix}


In Table~\ref{tab:extended}, we display additional information about the 15 civil enforcement actions highlighted in Table~\ref{tab:investgations}.

In Table~\ref{tab:terms}, we describe the search terms we used to compile our list of civil enforcement actions. We used a range of broad and legal-specific search engines including Google, HeinOnline, Lexis, and WestLaw to locate relevant public documents and winnowed according to the process described in our methodology. 

\clearpage
\onecolumn

\begin{small}
\begin{landscape}
    \begin{longtable}{p{3cm}p{2cm}p{3cm}p{3cm}p{4cm}p{3cm}}
\toprule
\textbf{Complainant(s)} & \textbf{Respondent(s)} & \textbf{Type of Action, Venue, Year} & \textbf{Regulation/Authority} & \textbf{Algorithm and Harm Alleged} & \textbf{Status (May 2025)} \\
\midrule

Louis and other rental applicants in Massachusetts & SafeRent Solutions LLC & Federal class action lawsuit, District of Massachusetts, 2022 & Federal Fair Housing Act (42 U.S.C. §§ 3604 et seq), Massachusetts antidiscrimination and consumer protection laws & Algorithmic tenant screening program alleged to have a disparate impact on low-income Black and Hispanic housing voucher recipients and housing applicants & Settlement reached \\
\midrule

US Federal Trade Commission & Rite Aid Corp. & Federal agency enforcement, Eastern District of Pennsylvania, 2023 & Section 5 of the FTC Act (15 U.S.C. §§ 45(a),(n)), prohibiting unfair acts or practices & Facial recognition technology alleged to erroneously identify shoplifters, with high false-positive matches especially among Black, Latino, Asian, and female consumers & Settlement reached \\
\midrule

US Equal Employment Opportunity Commission & iTutor & Federal agency enforcement, Eastern District of New York, 2022 & Federal Age Discrimination in Employment Act (ADEA) & Automated system in recruiting software alleged to reject female applicants over 55 years old and male applicants over 60 years old & Settlement reached \\
\midrule

US Department of Justice & Meta & Federal agency enforcement, Southern District of New York, 2022 & Federal Fair Housing Act (42 U.S.C. §§ 3601 et seq) & Ad-delivery system for housing advertisements alleged to discriminatorily target groups based on race, ethnicity, and sex & Settlement reached \\
\midrule

DeHoyos and other housing applicants in Texas and Florida & Allstate & Federal class action lawsuit, Western District of Texas, 2001 & Federal civil-rights claims (42 U.S.C. §§ 1981, 1982), Fair Housing Act & Automated credit scoring algorithms alleged to have adverse disparate impact on minority insurance applicants by charging unfairly high premiums & Settlement reached \\
\midrule

Huskey and other homeowners in the Midwest & State Farm & Federal class action lawsuit, Northern District of Illinois, 2022 & Federal Fair Housing Act (42 U.S.C. §§ 3604(a), (b), 3605) & Homeowner insurance claims-processing algorithms alleged to predict higher fraud levels from Black policyholders, subjecting them to greater scrutiny & Case is pending \\
\midrule

Mobley and other job applicants & Workday & Federal class action lawsuit, Northern District of California, 2023 & Title VII, ADEA, ADA, 42 U.S.C. § 1981, California Fair Employment and Housing Act & Job application automated screening tool alleged to discriminate against applicants based on race, age, and/or disability & Case is pending \\
\midrule

Flores and other inmates in New York & Stanford and other members of New York State Board of Parole & Federal class action lawsuit, Southern District of New York, 2018 & 8th Amendment, 14th Amendment, 42 U.S.C. § 1983 & Predictive risk assessment tool (COMPAS) alleged  to calculate recidivism scores without factoring demonstrated maturity and rehabilitation, minimizing younger inmates' chances of release & Settlement reached \\
\midrule

Oliver and other mortgage applicants & Navy Federal Credit Union & Federal class action lawsuit, Eastern District of Virginia, 2023 & Fair Housing Act, Equal Credit Opportunity Act, 42 U.S.C. § 1981 & Mortgage lending algorithm decisions alleged to discriminate based on race by denying loans or offering less favorable terms & Settlement reached \\
\midrule

Real Women in Trucking & Meta & EEOC complaint, Washington Field Office, 2022 & Title VII, ADEA & Ad-delivery algorithm alleged to target job applicants based on gender and age & Complaint pending \\
\midrule

Connecticut Fair Housing Center and Carmen Arroyo & CoreLogic Rental Property Solutions & Federal lawsuit, District of Connecticut, 2018 & Fair Housing Act, Connecticut Unfair Trade Practice Act, Fair Credit Reporting Act & Tenant screening algorithmic tool (CrimSAFE) alleged to discriminate against persons with criminal records, disproportionately impacting Black and Latino applicants & Appeal pending before 2nd Circuit \\
\midrule

National Fair Housing Alliance; Fair Housing Rights Center SE PA; Housing Equality Center PA & Tenant Turner & HUD complaint, 2024 & Fair Housing Act (42 U.S.C. § 3604(a), (b), (c)) & House screening algorithm alleged to discriminate based on race by prohibiting voucher holders from scheduling rental unit viewings & Complaint pending \\
\midrule

Liapes & Facebook & Class action lawsuit, California state court, 2020 & Unruh Civil Rights Act (Civ. Code, §§ 51, 51.5, 52) & Online advertising algorithm alleged to discriminate against women in showing life insurance ads & Case closed \\
\midrule

Buj & Psychiatry Residency Training & Federal lawsuit, District of New Jersey, 2017 & Title VII, ADA, ADEA, New Jersey Law Against Discrimination & Residency match algorithm alleged to discriminate against plaintiff based on national origin, religion, age, or disability & Case dismissed \\
\midrule

New York State Dept of Financial Services & Apple & NY Consumer Protection and Financial Enforcement Division Investigation, 2019 & Equal Credit Opportunity Act, New York State Human Rights Law & Credit-decision and credit-limit algorithms alleged to discriminate against women & No evidence of deliberate discrimination but deficiencies in customer service and transparency \\

\bottomrule \\ 
    \caption{Full list of algorithmic discrimination actions in our study.}
    \label{tab:extended}
    \end{longtable}
    \end{landscape}
\end{small}

\clearpage
\twocolumn

\begin{table*}[h]
\begin{tabular}{p{0.3\textwidth}p{0.5\textwidth}}
\toprule
\textbf{Category} & \textbf{Terms} \\
\midrule
Law Enforcement Investigation & ``Department of Justice investigation'' \\
& ``FTC investigation'' \\
& ``State Attorney General investigation'' \\
& ``regulatory probe'' \\
& ``civil rights investigation'' \\
\midrule
Algorithmic Decision-Making & ``algorithmic bias'' \\
& ``algorithmic discrimination'' \\
& ``automated decision system'' \\
& ``predictive algorithm'' \\
& ``AI-based decisions'' \\
& ``automated screening'' \\
& ``scoring system'' \\
& ``risk assessment tool'' \\
\midrule
Public Legal Documentation & ``findings report'' \\
& ``consent decree'' \\
& ``settlement agreement'' \\
& ``press release'' \\
& ``public statement'' \\
\bottomrule
\end{tabular}
\caption{Search terms used for compiling civil enforcement actions}
\label{tab:terms}
\end{table*}

\end{document}